# Iron oxide nanoparticles for magnetic hyperthermia

**N. A. Usov[1,2]**

[1]*National University of Science and Technology «MISIS», 119049, Moscow, Russia*
[2]*Pushkov Institute of Terrestrial Magnetism, Ionosphere and Radio Wave Propagation, Russian Academy of Sciences, (IZMIRAN) 108480, Troitsk, Moscow, Russia*

**Abstract**

Assemblies of magnetic nanoparticles show a great potential for application in biomedicine, in particular, in magnetic hyperthermia. However, to achieve desired therapeutic effect in magnetic hyperthermia the assembly of nanoparticles should have a sufficiently high specific absorption rate (SAR) in alternating magnetic field of moderate amplitude and frequency. Using the Landau – Lifshitz stochastic equation it is shown that dilute assemblies of iron oxides nanoparticles of optimal diameters are capable of providing SAR of the order of 400 – 600 W/g in alternating magnetic field with the amplitude $H_0$ = 100 Oe, in the frequency range $f$ = 300 – 500 kHz. Unfortunately, in dense clusters of magnetic nanoparticles, which are often formed in a biological medium, there is a sharp decrease in SAR due to the influence of strong magneto- dipole interaction of closest nanoparticles. To overcome this difficulty it is suggested covering the nanoparticles with nonmagnetic shells of sufficient thickness, or using non single-domain nanoparticles being in magnetization curling states.

## 1. Introduction

Magnetic nanoparticles are widely used in various areas of nanotechnology, such as high density magnetic recording, manufacture of permanent magnets, etc. [1,2]. Recently, assemblies of magnetic nanoparticles have been actively studied for application in biomedicine, for targeted drug delivery, purification of biological media, in magnetic hyperthermia, and other promising areas [3].

The method of magnetic hyperthermia [3-7] is based on the absorption of an alternating magnetic field energy by magnetic nanoparticles distributed in a biological medium. The local remote heating of the malignant tissues in a living organism in most cases leads to stunting and destruction of cancer tumors. But to achieve a positive therapeutic effect it is necessary to heat the tumor tissue to temperatures of about 42–45° C, possibly without exposing the surrounding healthy tissue to excessive heat.

The experimental and theoretical results obtained so far [4-27] show a great potential of magnetic hyperthermia. However, a number of difficulties still need to be overcome for the widespread introduction of this technique into clinical practice. In fact, an assembly of magnetic nanoparticles is a very complicated physical object. The magnetic properties of the assembly are determined by a variety of geometrical and physical factors, such as the complex nature of the magnetic anisotropy of individual nanoparticles, the influence of thermal fluctuations of particle magnetic moments, the action of a strong magneto- dipole interaction between nanoparticles of a dense assembly.

To achieve desired therapeutic effect in magnetic hyperthermia an assembly of nanoparticles should have a sufficiently high specific absorption rate (SAR) in alternating magnetic field of moderate amplitude and frequency. In addition, for use in biomedical applications the magnetic nanoparticles must be weakly toxic for a living organism, and simultaneously be resistant to the aggressive effects of the biological environment [5-7].

It was recently realized [9] that commercially produced assemblies of magnetic nanoparticles are unlikely to satisfy the entire complex of the above conditions. Significant efforts are required to develop advanced methods of synthesis of magnetic nanoparticle assemblies specially optimized for use in magnetic hyperthermia. A detailed theoretical analysis [4,11,20,27] helps to determine the conditions for assembly to effectively absorb the energy of alternating magnetic field.

This paper is devoted to the discussion of some theoretical issues associated with the use of magnetic nanoparticle assemblies in magnetic hyperthermia. A short review of the properties of superparamagnetic nanoparticles is also given for completeness.

## 2. Properties of superparamagnetic nanoparticles

Many ferromagnetic materials are able to absorb the energy of an external alternating magnetic field. However, magnetic nanoparticles have significant advantages for use in magnetic hyperthermia: a) assemblies of superparamagnetic nanoparticles are capable of providing extremely large SAR values, of the order of 1 kW per gram of substance; b) due to their small size, nanoparticles can penetrate deeply into biological materials; c) iron oxide nanoparticles are non-toxic, or slightly toxic for a living organism, d) they have short periods of elimination from the body. Due to the low toxicity and sufficiently high saturation magnetization, iron oxide nanoparticles and some ferrites are considered most suitable for biomedical applications [2-7].

Magnetic nanoparticles of iron oxides are characterized by saturation magnetization $M_s$ = 350 - 450 emu/cm$^3$, exchange constant $C = 2\times10^{-6}$ erg/cm, and effective magnetic anisotropy constant $K = 10^4 - 10^5$ erg/cm$^3$ [28]. It is well known [29, 30] that in a ferromagnetic sample the magnitude of the magnetization vector $\vec{M}(\vec{r})$ is equal to the saturation magnetization, $|\vec{M}(\vec{r})| = M_s$, whereas the direction of this vector is described by a unit magnetization vector, $\vec{\alpha}(\vec{r}) = \vec{M}(\vec{r})/M_s$, $|\vec{\alpha}(\vec{r})| = 1$. The exchange energy of the uniformly magnetized nanoparticle, $\vec{\alpha}(\vec{r}) = \vec{\alpha}_0$, is zero, $w_{exc} = 0$. In fact, the uniformly magnetized particle is a small permanent magnet that creates a strong magnetic field around and has magnetic energy per unit volume on the order of $w_m \sim M_s^2$. The magnetic energy is significantly reduced for a particle being in magnetization curling state. Simultaneously, the exchange energy of the nanoparticle of characteristic radius $R$ will increase to a value of the order of $w_{exc} \sim C/R^2$. For a magnetically soft nanoparticle ($K < M_s^2$), the homogeneous state will have the lowest total energy provided that $C/R^2 \geq M_s^2$. Equating these values, one can estimate qualitatively the single-domain radius of a magnetically soft nanoparticle, $R_c \sim C^{1/2}/M_s \sim 25 - 30$ nm.

In most cases, single-domain magnetic nanoparticles with sizes $D < 2R_c$ are most interesting for applications. In the exact calculation of the single-domain radius $R_c$ one has to take into account the magnetic anisotropy energy. The latter is usually a small correction to the sum of the exchange and magnetic energies of a single-domain nanoparticle. But it is precisely the energy of magnetic anisotropy that determines the stationary directions of the unit magnetization vector [28–31]. Indeed, since the exchange energy of uniform magnetization is zero, it does not depend on the direction of the unit magnetization vector. Although the magnetic energy of a uniformly magnetized particle is large, $w_m \sim M_s^2$, for a spherical nanoparticle the magnetic energy also does not depend on the direction of the magnetization vector due to symmetry in the distribution of magnetic charges on the surface of a sphere. At the same time, in the absence of an external magnetic field the unit magnetization vector of a single-domain spherical nanoparticle is oriented in strictly defined directions with respect to the axes of symmetry of the particle crystal lattice. These distinguished spatial directions are called the easy anisotropy axes of the nanoparticle. They are determined by the equation for the energy density of magneto-crystalline anisotropy, $w_a = w_a(\vec{\alpha})$.

It follows from general considerations [29-31] that for crystals with a single axis of symmetry the energy density of magneto-crystalline anisotropy can be represented as an expansion in powers of the unit magnetization vector

$$w_a(\vec{\alpha}) = K_1(\alpha_x^2 + \alpha_y^2) + K_2(\alpha_x^2 + \alpha_y^2)^2 + ... \quad (1)$$

where it is assumed that the axis of symmetry is parallel to the z axis of the Cartesian coordinates. In Eq. (1), $K_1$, $K_2$, etc are uniaxial magnetic anisotropy constants, which usually decrease in absolute value, $|K_1| > |K_2|$. It follows from Eq. (1) that in the case $K_1$, $K_2 > 0$ the energy density of uniaxial magnetic anisotropy is minimal, $w_a = 0$, if the unit magnetization vector is parallel to the z axis, that is, $\vec{\alpha} = (0,0,1)$ or $\vec{\alpha} = (0,0,-1)$. Thus, for a particle with anisotropy energy density, Eq. (1), these directions of the unit magnetization vector are preferable.

On the other hand, for crystal lattice of a cubic symmetry the energy density of the magneto-crystalline anisotropy can be written as follows [29-31]

$$w_a(\vec{\alpha}) = K_{1c}(\alpha_x^2\alpha_y^2 + \alpha_x^2\alpha_z^2 + \alpha_y^2\alpha_z^2) + K_{2c}\alpha_x^2\alpha_y^2\alpha_z^2 + ... \quad (2)$$

In the case $K_{1c} > 0$, $K_{2c} \approx 0$, the directions of the easy anisotropy axes of a particle with cubic anisotropy are parallel to the axes of the Cartesian coordinates, since the magnetic anisotropy energy, Eq. (2), is minimal, $w_a = 0$, when the unit magnetization vector is parallel to the $x$, $y$ or $z$ axis, i. e. $\vec{\alpha} = (\pm 1,0,0)$, etc. Thus, in the case $K_{1c} > 0$ the particle has 6 equivalent directions of the easy anisotropy axis. If the constant $K_{1c} < 0$, the directions of easy anisotropy axis are parallel to the cube diagonals, since in this case the magnetic anisotropy energy density, Eq. (2), has minima for the vectors $\vec{\alpha} = (\pm 1/\sqrt{3}, \pm 1/\sqrt{3}, \pm 1/\sqrt{3})$.

It should be noted, however, that Eqs. (1), (2) are only the simplest energy contributions that determine the spatial directions of the nanoparticle easy anisotropy axes. Another important contribution to the effective magnetic anisotropy is related with a deviation of particle shape from sphere. This contribution is called shape anisotropy energy [29,30]. It is especially important for nanoparticles of the soft magnetic type, with a fairly high saturation magnetization. The famous Brown – Morrish theorem [32] states that the magnetostatic energy of a uniformly magnetized particle of arbitrary shape in the first approximation coincides with that of some equivalent ellipsoid. If the Cartesian coordinate axes are chosen along the symmetry axes of an equivalent ellipsoid, the magnetic energy density of the nanoparticle has the form [32]

$$w_m(\vec{\alpha}) = \frac{1}{2}M_s^2(N_x\alpha_x^2 + N_y\alpha_y^2 + N_z\alpha_z^2), \quad (3)$$

where $N_x$, $N_y$, $N_z$ are the demagnetizing factors of the equivalent ellipsoid in the given coordinate system.

Obviously, by virtue of symmetry the demagnetizing factors of a uniformly magnetized cube are equal to each other, $N_x = N_y = N_z = 4\pi/3$. Therefore, it follows from the condition $\alpha_x^2 + \alpha_y^2 + \alpha_z^2 = 1$ that the magnetic energy of a cube, like the magnetic energy of a sphere, does not depend on the direction of the unit magnetization vector and does not contribute to the magnetic anisotropy of such particle. On the other hand, if the particle has the shape of an elongated spheroid, whose transverse demagnetizing factors are equal, $N_x = N_y > N_z$, the magnetic energy of the particle, Eq. (3), can be written as

$$w_m(\vec{\alpha}) = \frac{1}{2}M_s^2(N_x - N_z)(\alpha_x^2 + \alpha_y^2) + const \qquad (4)$$

Comparing Eqs. (1) and (4), one can conclude that an elongated equivalent spheroid can be characterized by an effective shape anisotropy constant, $K_{ef} = M_s^2(N_x - N_z)/2$. However, both the direction of the easy anisotropy axis and the values of the effective demagnetizing factors $N_x$ and $N_z$ are determined [10] by the actual shape of the nanoparticle. In the general case the total magnetic anisotropy energy is the sum of the magneto-crystalline and the shape anisotropy energies, respectively. Such nanoparticle has a combined magnetic anisotropy [33].

To clearly demonstrate the arrangement of the minima and maxima of the magnetic anisotropy energy, it is convenient to write this energy in spherical coordinates, $w_a = w_a(\theta, \varphi)$, [28]. To do this it is sufficient to express in Eqs. (1) - (4) the components of the unit magnetization vector in terms of spherical angles ($\theta$, $\varphi$). Then one can build a surface $r(\theta, \varphi) = w_a(\theta, \varphi)$. Fig. 1a shows the reduced energy density of the uniaxial magnetic anisotropy $w_a(\theta, \varphi)/K_1$, Eq. (1), for the case $K_1 > 0$, $K_2 = 0$. Obviously, the magnetic anisotropy energy has deep minima for the directions of the unit magnetization vector, close to the positive or negative direction of the $z$ axis, i. e. to the angles $\theta = 0, \pi$. These minima are separated by a potential barrier whose maximum corresponds to the angle $\theta = \pi/2$. Fig. 1b shows the reduced energy density of cubic magnetic anisotropy, $w_a(\theta, \varphi)/K_{1c}$, Eq. (2), for the case of $K_{1c} > 0$, $K_{2c} = 0$. In this case, the minima of the anisotropy energy correspond to directions parallel to the axes of the Cartesian coordinates. In total, there are 6 energy minima separated by barriers. At the same time for the case $K_{1c} < 0$, $K_{2c} = 0$, shown in Fig. 1c, energy minima correspond to the directions parallel to the cube diagonals. Consequently, there are 8 equivalent directions for which the particle anisotropy energy has a minimum.

The energy surfaces shown in Figs. 1a to 1c correspond to an ideal case of a spherical magnetic nanoparticle. If the shape of a particle deviates from sphere, the shape anisotropy energy, Eq. (4), has to be added to the magneto-crystalline anisotropy energy, Eq. (1) - (3). The shape anisotropy energy may give a significant contribution to the total anisotropy energy of a particle with a large saturation magnetization. As an example, Fig. 1d shows the case of particle with combined magnetic anisotropy. Here, the reduced shape anisotropy energy, $K_{ef}\left(1-(\vec{\alpha}\vec{n})^2\right)/K_{1c}$, with the anisotropy constants ratio $K_{ef}/K_{1c} = 0.5$ is added to the energy density of the cubic anisotropy shown in Fig. 1b. The unit vector ***n*** shows the direction of the easy axis of shape anisotropy, which in this particular case is given by spherical angles $\theta = \pi/6$, $\varphi = \pi/3$.

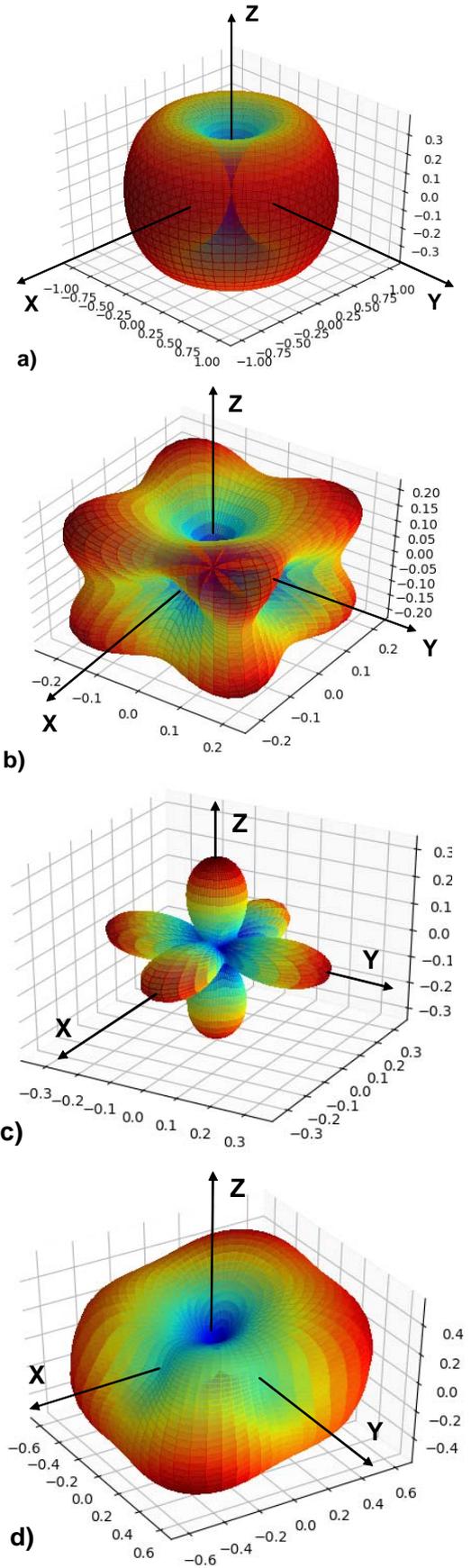

Fig.1. Reduced magnetic anisotropy energy density, $w_a = w_a(\theta, \varphi)/K$, for various cases: a) uniaxial magnetic anisotropy with easy axis parallel to the $z$ axis, b) cubic anisotropy in the case of $K_{1c} > 0$, c) cubic anisotropy in the case of $K_{1c} < 0$, d) the case of a combined magnetic anisotropy.

Comparing Fig. 1b and 1d, it is easy to see that due to the influence of the shape anisotropy energy the surface of the combined anisotropy energy becomes two-pit, but the barrier separating the potential wells may have a very complex shape. It is clear that the shape of the energy surface of the combined magnetic anisotropy substantially depends on the ratio $K_{ef}/K_{1c}$ and the direction of the vector $n$ with respect to the axes of the cubic magnetic anisotropy.

It was shown [32] that for nanoparticles of iron oxides a relatively small spheroidal distortion of a nanoparticle in a certain direction leads to a significant contribution of the shape anisotropy energy. Therefore, it can be expected that the situation shown in Fig. 1d is fairly typical. Because of this, in most cases it is assumed that the nanoparticles of iron oxides in real experimental assemblies have effective uniaxial anisotropy.

As emphasized above, single-domain nanoparticles are most interesting for applications, since they have a permanent total magnetic moment. To accurately determine the single-domain radius of a magnetic nanoparticle, it is necessary to compare the total energy of uniform magnetization with the energy of the lowest inhomogeneous micromagnetic state. It is easy to show [29, 30] that the total energy of uniform magnetization does not depend on the nanoparticle size. For magnetically soft particles a vortex competes in energy with the uniform magnetization [29, 30].

The structure and energy of the vortex state can be calculated using numerical simulation [34,35]. Fig. 2 shows the energy diagram of stationary states in a spherical magnetite nanoparticle with the saturation magnetization $M_s$ = 450 emu/см$^3$, the exchange constant $C = 2 \times 10^{-6}$ erg/cm, and the cubic magnetic anisotropy constant $K_{1c} = -10^5$ erg/cm$^3$ [28]. As Fig. 2 shows, the total energy of uniform magnetization, curve (1), does not depend on the particle diameter, while the total energy of the vortex, curve (2), decreases sharply as a function of particle diameter. The single-domain diameter of a spherical magnetite nanoparticle, $D_c$ = 64 nm, is determined by the intersection of the corresponding energy curves 1) and 2) in Fig. 2.

As discussed above, taking into account the magnetic anisotropy energy, the magnetic moment of the nanoparticle has several stable spatial directions for which the total nanoparticle energy has local minima. At zero temperature the magnetic moment of a single-domain nanoparticle is located in one of the energy minima. At room temperature the magnetic moment of the nanoparticle performs an irregular precession of small amplitude near the energy minimum under the influence of thermal fluctuations.

In an applied external magnetic field, the energy minima of the potential wells shift relative to each other, some minima disappear, others appear. The magnetic moment of a particle can jump from one potential well to another when the magnitude or direction of external magnetic field is changed. The dynamics of a unit magnetic vector of a nanoparticle in external quasistatic or alternating magnetic field $H_0(t)$ is described by the Landau – Lifshitz equation [36].

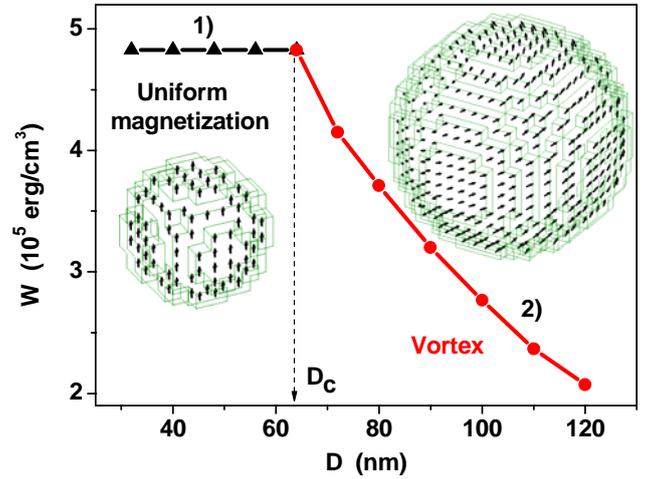

Fig. 2. The total energy of a uniformly magnetized 1) and vortex 2) states in a spherical magnetite nanoparticle depending on its diameter.

The latter, along with the external magnetic field, takes into account the exchange and anisotropic interactions, as well as the demagnetizing field created by magnetic charges distributed in the bulk and on the surface of the nanoparticle.

Note that a single-domain nanoparticle is a strong natural magnet, since the characteristic field of the particle magnetization reversal at room temperature can be rather large, $H_c \sim 2K/M_s \sim 400$ Oe. However, the height of the energy barrier separating the magnetic potential wells, $\Delta E \sim KV$, decreases with a decrease in the particle volume $V$. For small particle sizes the magnetic anisotropy energy can be compared with the characteristic thermal energy $k_B T$, where $k_B$ is the Boltzmann constant. The characteristic residence time of the magnetic moment in a given potential well (the Neél relaxation time) is estimated to be $\tau_N = \tau_0 \exp(KV/k_B T)$, where the constant $\tau_0 \sim 10^{-9} - 10^{-11}$ sec [37-40]. Therefore, the relaxation time decreases exponentially rapidly with decreasing particle diameter. As soon as $\tau_N$ becomes about or less than the characteristic time of the magnetic moment measuring, the average magnetic moment of the nanoparticle turns out to be zero. Thus, thermal fluctuations may significantly limit the time of stable storage of information in the magnetic recording technique [40]. But the phenomenon of superparamagnetism is helpful for magnetic hyperthermia. Thermal fluctuations, swinging the magnetic moment in a potential well, effectively lower the energy barrier and significantly reduce the characteristic field for particle magnetization reversal. As a result, the magnetization reversal of an assembly of superparamagnetic particles is possible in alternating magnetic field of moderate amplitude, $H_0 \sim 100 - 200$ Oe.

For theoretical description of the influence of thermal fluctuations on the particle behavior W.F. Brown [38] suggested to use the Landau – Lifshitz stochastic equation (see Eqs. (5), (6) below). In this equation, the influence of thermal fluctuations on the dynamics of the particle magnetic moment is described by the action of a random magnetic field with the statistical characteristics of white noise. It is convenient to implement this approach by means of numerical simulation [41-44].

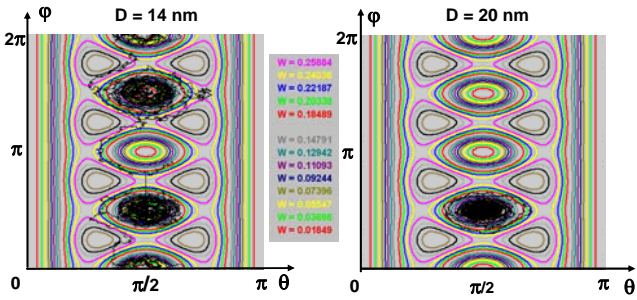

Fig. 3. Dynamics of the unit magnetization vectors of iron nanoparticles of various diameters on the surface of a unit sphere under the action of thermal fluctuations at $T$ = 300 K.

Fig. 3 shows the temporal dynamics of the unit magnetization vector of a spherical iron nanoparticle on the surface of a unit sphere, $0 \leq \theta \leq \pi$, $0 \leq \varphi \leq 2\pi$, at a room temperature, $T$ = 300 K. The saturation magnetization of the particle is $M_s$ = 1700 emu/cm$^3$, the cubic magnetic anisotropy constant of iron is positive, $K_{1c}$ = 5.5×10$^5$ erg/cm$^3$. In Fig. 3 the smooth curves of different colors show the level lines for the particle anisotropy energy density, Eq. (2). The irregular black curves show the trajectories of the unit magnetization vector for 10$^6$ numerical steps with a small time increment $\Delta t$ = 4×10$^{-14}$ sec. The left panel in Fig. 3 corresponds to the nanoparticle with diameter $D$ = 14 nm. For this nanoparticle the reduced energy barrier between the potential wells is small, $K_{1c}V/4k_BT$ = 4.8. As a result, under the influence of thermal fluctuations the unit magnetization vector of this particle easily jumps between various potential wells during the 10$^6$ time steps performed. For comparison, the right panel of Fig. 3 shows the case of iron nanoparticle with diameter $D$ = 20 nm. For this particle the reduced energy barrier between the wells is significantly higher, $K_{1c}V/4k_BT$ = 13.9. As a result, the unit magnetization vector remains in the same potential well during 10$^6$ iterations with the same time step.

Thus, one can see that the properties of assemblies of superparamagnetic nanoparticles are determined by a large set of geometric and magnetic parameters, namely, particle sizes, external shapes, distribution of easy anisotropy axes, saturation magnetization, ambient temperature, etc. It is also very important whether the particle is monocrystalline, or consists of different crystallites of various spatial orientations connected by exchange interaction. In real experimental assemblies there is usually a substantial scatter of nanoparticles in size and shape. In addition, the particles obtained by means of chemical synthesis are often polycrystalline. In a polycrystalline nanoparticle anisotropic interactions are substantially averaged, which affects the characteristic single-domain size and coercive force of the particle [45]. These circumstances significantly complicate the interpretation of experimental results. In addition, the magnetic properties of such assemblies are very difficult to control.

## 3. Heating ability of dilute assemblies of nanoparticles

Let us consider now the behavior of an assembly of superparamagnetic nanoparticles in an alternating external magnetic field under the assumption that the particles of the assembly are distributed in a dense biological medium, such as a tumor. In this case, the nanoparticles are tightly connected with the surrounding biological tissues, so that their movement and rotation as a whole are inhibited. Consequently, only the magnetic moments of the nanoparticles can respond to the action of a variable external magnetic field. In this section, we will also assume that the assembly of nanoparticles is sufficiently dilute, so that the effect of the magneto-dipole interaction on the assembly behavior can be neglected.

### 3.1. *Superparamagnetic nanoparticles*

It has been noted above that the coercive force of an assembly of superparamagnetic nanoparticles strongly depends on the characteristic particle diameter, as well as on the temperature. Because of this, in some cases at room temperature the magnetization reversal of an assembly of superparamagnetic nanoparticles is possible in alternating magnetic fields of relatively small amplitude, $H_0$ ~ 50 - 100 Oe. This is especially important in magnetic hyperthermia, since the creation and use of variable magnetic fields of large amplitude is expensive and may be unsafe in a medical clinic.

The use of assemblies of ferromagnetic nanoparticles in magnetic hyperthermia [3–7] is based on the fact that cyclical reversal of ferromagnetic particles generates specific heat per unit of time proportional to the product of the hysteresis loop area of the assembly and the magnetization reversal frequency. Thus, the amount of thermal energy released increases with increasing of the hysteresis loop area and frequency. However, according to the empirical Brezovich's criterion [6,46], the alternating magnetic field is harmless to the human body if its amplitude $H_0$ and frequency $f$ satisfy the condition $fH_0 < 5\times10^9$ A/(ms).

The calculation of low-frequency hysteresis loops of an assembly of superparamagnetic nanoparticles can be carried out using the Landau – Lifshitz stochastic equation [11]. This method requires significant CPU time. However, it is applicable to assemblies of nanoparticles with an arbitrary type of magnetic anisotropy. In addition, this approach is generalized [27] to the case of dense assemblies of magnetic nanoparticles with a strong magneto- dipole interaction between the nanoparticles.

The stochastic Landau-Lifshitz equation is given by [38, 41-44, 35]

$$\frac{d\vec{\alpha}}{dt} = -\gamma_1 \vec{\alpha} \times \left(\vec{H}_{ef} + \vec{H}_{th}\right) - \kappa\gamma_1 \vec{\alpha} \times \left(\vec{\alpha} \times \left(\vec{H}_{ef} + \vec{H}_{th}\right)\right) \quad (5)$$

where $\kappa$ is the magnetic damping parameter, $\gamma$ is the gyromagnetic ratio, $\gamma_1 = \gamma/(1+\kappa^2)$, $\vec{H}_{ef}$ is the effective magnetic field and $\vec{H}_{th}$ is the thermal field. The latter is assumed to be a Gaussian random process with the following statistical properties of its components [38]

$$\langle H_{th,i}(t)\rangle = 0\,;\, \langle H_{th,i}(t)H_{th,j}(t')\rangle = \frac{2k_B T\kappa}{M_s V\gamma}\delta_{i,j}\delta(t-t'), \quad (6)$$

where $i, j = (x,y,z)$. The numerical simulations are usually performed in the intermediate-to-high damping limit, $\kappa = 0.5 - 1.0$, due to the large number of structural defects in magnetic nanoparticles.

To ensure the accuracy of the numerical simulations it is necessary to keep the physical time step $\Delta t$ sufficiently small with respect to the characteristic particle precession time, $T_p \sim 1/\gamma H_a$, where $H_a = 2K/M_s$ is the particle anisotropy field. It can be shown [11] that in the case of intermediate-to-high damping limit it is enough to keep the ratio $\Delta t/T_p < 1/20$. Further reduction of this ratio only slightly changes the particle hysteresis loop. To get hysteresis loop of a dilute nanoparticle assembly it is sufficient to calculate time dependent magnetization $M(t)$ of an isolated ferromagnetic particle in a large series of statistically independent numerical experiments with fixed initial conditions. Then, the average assembly magnetization is given by

$$\langle \vec{M}(t)\rangle = \frac{1}{N_{exp}}\sum_{n=1}^{N_{exp}} \vec{M}_n(t). \quad (7)$$

It has been found empirically [11], that $N_{exp} \sim 1000 - 2000$ is usually sufficient to reduce the random oscillations of the average assembly magnetization up to several percent.

To eliminate the influence of the initial conditions which can be arbitrarily specified, a sufficiently large number of time steps are performed in each trial, $N_{step} \gg T_H/\Delta t$, where $T_H = 1/f$ is the period of the alternating magnetic field. It can be proved that one obtains a steady assembly hysteresis loop extending the calculation over a sufficiently large number of periods $T_H$.

Fig. 4 shows the calculated low-frequency hysteresis loops and the SAR for a random assembly of non-interacting superparamagnetic nanoparticles with uniaxial anisotropy. Here, typical values of saturation magnetization $M_s = 350$ emu/cm$^3$, and the effective magnetic anisotropy constant $K = 10^5$ erg/cm$^3$, are adopted for iron oxide nanoparticles. The effective uniaxial magnetic anisotropy observed experimentally is apparently due to the influence of particle shape anisotropy. The calculations shown in Fig. 4 are carried out in alternating magnetic field of a fixed amplitude, $H_0 = 100$ Oe, in the frequency range $f = 200 - 500$ kHz. The assembly temperature is $T = 300$ K, the magnetic damping constant is taken to be $\kappa = 0.5$.

As Fig. 4a shows, there is a sharp dependence of the low-frequency hysteresis loops area on the particle diameter. This behavior can be interpreted based on simple qualitative considerations [11]. The main conclusion is that the hysteresis loop area of a superparamagnetic assembly is nonzero under the condition $f\tau \sim 1$, where $\tau$ is the characteristic relaxation time of the particle magnetic moment.

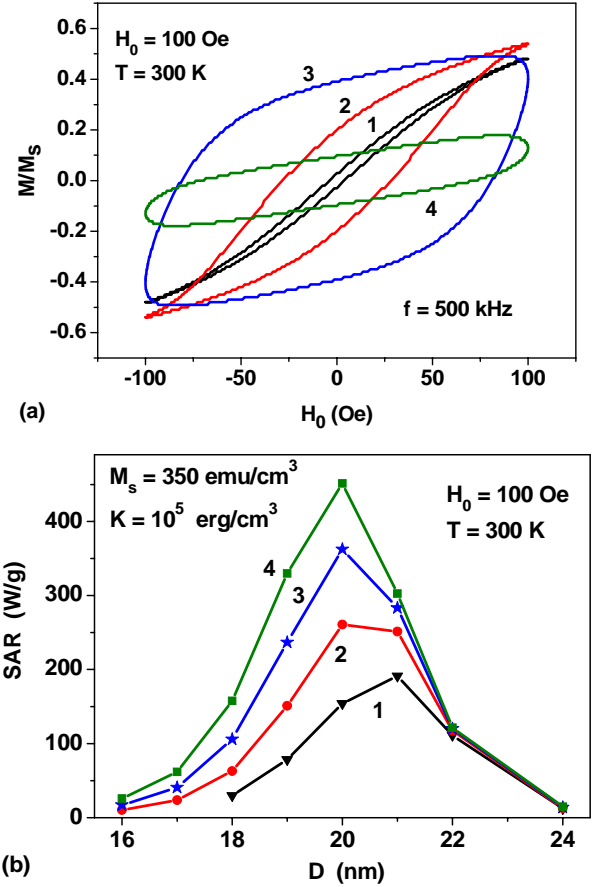

Fig. 4. a) Hysteresis loops of an assembly of non-interacting nanoparticles with uniaxial anisotropy of different diameters at a frequency $f = 500$ kHz: 1) $D = 16$ nm, 2) $D = 18$ nm, 3) $D = 20$ nm, 4) $D = 22$ nm; b) specific absorption rate of the assembly depending on the particle diameter at fixed magnetic field amplitude $H_0 = 100$ Oe and various frequencies: 1) $f = 200$ kHz, 2) $f = 300$ kHz, 3) $f = 400$ kHz, 4) $f = 500$ kHz.

Under the condition $f\tau \ll 1$ the assembly magnetic moment has a time to adjust to the change in the external magnetic field. In this case the magnetic hysteresis is small, since the magnetic moment of the assembly evolves along the equilibrium magnetization curve. This behavior is exemplified in Fig. 4a by the curve 1, which corresponds to the assembly of nanoparticles with diameter $D = 16$ nm. In the other limiting case, $f\tau \gg 1$, the relaxation times are large; therefore, the magnetization reversal of nanoparticles is unlikely during the magnetic field period. As a result, the area of the assembly hysteresis loop decreases significantly. This behavior is shown by the curve 4 in Fig. 4a, which corresponds to an assembly of nanoparticles with diameter $D = 24$ nm. If one takes into account that in the case of small field amplitudes, $H_0 < H_c$, to the order of magnitude $\tau \sim \exp(KV/k_B T)$ [37-38], then the sharp dependence of the hysteresis loop area on the particle diameter becomes clear.

It is well-known [31], that the thermal rate released per unit particle volume is determined by the integral

$$P = M_s f\oint \vec{\alpha}\, d\vec{H} = fA, \quad (8)$$

where $A$ is the hysteresis loop area in the variables ($M$, $H$). SAR per unit mass of the assembly is then given by SAR = $P/\rho$, [11] . The SAR value is measured in W/g if one measures a volumetric rate $P$ dissipated in ferromagnetic nanoparticles in W/cm$^3$, and a particle density in g/cm$^3$. For iron oxide nanoparticles the density $\rho = 5.0$ g/cm$^3$ is usually accepted.

Fig. 4b shows the dependence of the SAR on the particle diameter at various frequencies of alternating magnetic field. As can be seen, the SAR reaches a maximum for an assembly with the largest hysteresis loop area. In addition, in accordance with Eq. (8), SAR increases with increasing frequency. But the frequency dependence of the SAR is in fact non-linear, since the hysteresis loop area is also frequency dependent, $A = A(f)$. At sufficiently high frequencies the area of the hysteresis loop may decrease [11].

Similar calculations of the low-frequency hysteresis loops were also carried out for spherical magnetite nanoparticles with cubic anisotropy. As Fig. 5 shows, for this assembly at small magnetic field amplitudes, $H_0 = 50$ Oe, the SAR maximum shifts to larger particle diameters, $D = 45-50$ nm. This is due to the relatively small value of the energy barriers for particles with cubic anisotropy. With increase in field amplitude, $H_0 = 100$ Oe, the SAR increases as a function of diameter up to the single-domain particle diameter $D_c = 64$ nm. Note also that rather large values of SAR ~ 400 - 600 W/g, shown in Fig. 5, are obtained for dilute assemblies of nanoparticles neglecting the influence of magneto - dipole interaction.

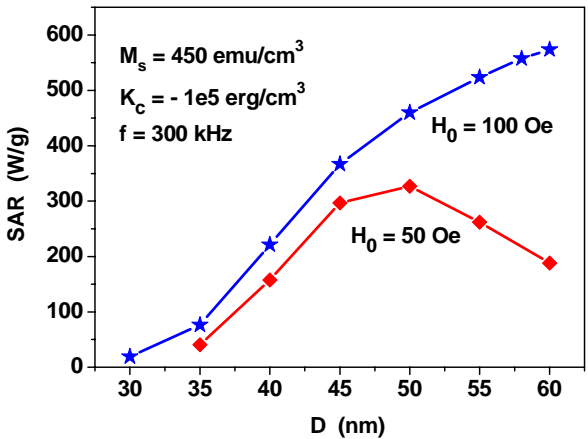

Fig. 5. SAR of a randomly oriented non interacting assembly of magnetite nanoparticles with cubic magnetic anisotropy as a function of particle diameter.

### 3.2. *Magnetic vortices*

Due to the low coercive force of an assembly of superparamagnetic nanoparticles, most theoretical and experimental studies in magnetic hyperthermia are carried out for assemblies of nanoparticles with diameters substantially smaller than the single-domain one. However, it was recently shown [47] that an assembly of magnetically soft nanoparticles with diameters greater than the single-domain diameter can also effectively absorb the energy of alternating magnetic field. As Fig. 2 shows, the particles of such an assembly are in vortex states. The average magnetic moment of the vortex remains rather large, $<M>/M_s = 0.5–0.8$, if the diameter of the nanoparticle does not exceed too much the single-domain diameter. However, the coercive force of the vortex decreases, so that the magnetization reversal is possible in alternating magnetic fields of moderate amplitude, $H_0 = 100 - 150$ Oe. For quasi-spherical magnetite nanoparticles in vortex states, it was shown [47] that the range of particle diameters $D = 70 - 80$ nm is optimal for application in magnetic hyperthermia.

To calculate the SAR of a randomly oriented assembly of non single-domain nanoparticles with cubic anisotropy, it is necessary to obtain partial hysteresis loops of isolated nanoparticle for directions of alternating magnetic field lying in the domain of the irreducible directions shown in Fig. 6a. Fig. 6b shows the calculated partial hysteresis loops of magnetite nanoparticle with diameter $D = 72$ nm for the characteristic directions of the magnetic field 1) – 4) in Fig. 6a. As Fig. 6b shows, at a frequency $f = 1$ MHz and magnetic field amplitude $H_0 = 100$ Oe, all the particular hysteresis loops in the irreducible domain of directions have a sufficiently large area. By averaging the partial hysteresis loops over the irreducible region, the SAR of a randomly oriented assembly of nanoparticles with a diameter $D = 72$ nm at $f = 1$ MHz, $H_0 = 100$ Oe was estimated [47] to be SAR = 850 W/g. However, the SAR decreases to 460 W/g when the particle diameter increases up to $D = 80$ nm.

It is also important to note that in the case of non single-domain nanoparticles with diameters $D > D_c$ the volume of the heat generation is $(D/d)^3$ times larger than that for small superparamagnetic nanoparticle of diameter $d < D_c$.

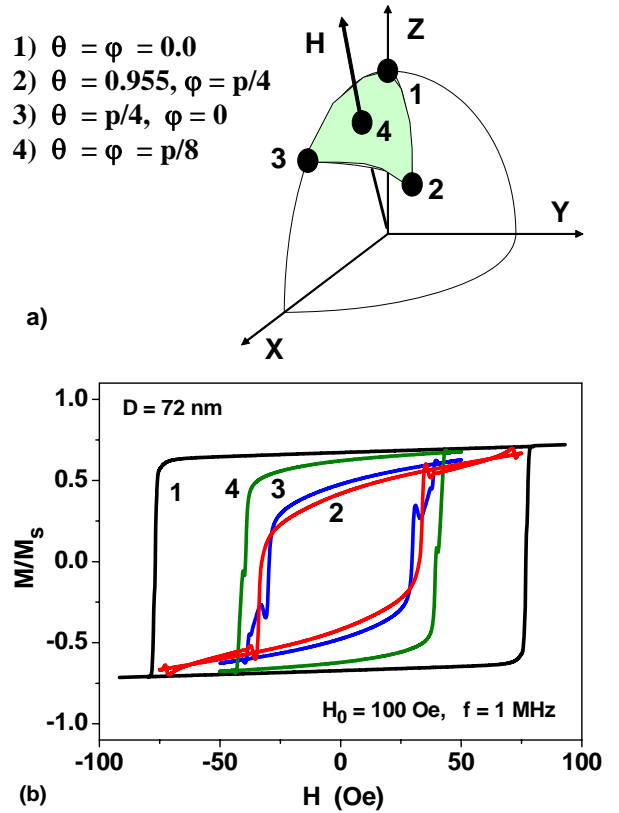

Fig. 6. a) Characteristic directions of applied magnetic field for a nanoparticle with cubic magnetic anisotropy; b) particular hysteresis loops of magnetite nanoparticle with diameter $D = 72$ nm for the directions 1) – 4) of applied alternating magnetic field shown in Fig. 6a).

## 4. Magneto-dipole interaction in a dense nanoparticle assembly

Recently, it has been experimentally proved [19,25,26] that magnetic nanoparticles introduced into a biological medium are subject to agglomeration. Penetrating into biological cells, or being in the intercellular space, they form dense clusters of fractal geometrical structure. In contrast to the usual cluster of nanoparticles, whose structure is shown in Fig. 7a, the structure of fractal cluster [48,49] is characterized by fractal descriptors $D_f$ and $k_f$. Evidently, the number of particles in an ordinary three-dimensional cluster of a fixed nanoparticle density is proportional to the cluster volume, i.e. to the cube of the characteristic cluster radius, $N_p \sim R^3$. However, for a fractal cluster consisting of particles of diameter $D$, the dependence of the number of particles on the characteristic cluster size $R_g$ is given by [49]

$$N_p = k_f (2R_g/D)^{D_f}, \qquad (9)$$

where $D_f$ is the fractal dimension, $k_f$ being the fractal prefactor. The radius of gyration $R_g$ of fractal cluster is defined as the mean square of the distances of cluster particles from the geometric center of mass of the cluster $\vec{R}_0$,

$$R_g = \left( \sum_i (\vec{r}_i - \vec{R}_0)^2 / N_p \right)^{1/2}.$$

It is important to note that for a cluster of fractal structure, the fractal dimension can be fractional. As an example, Fig. 7b shows the geometric structure of the cluster with fractal descriptors $D_f = 1.9$, $k_f = 1.7$. Clusters of this type are often formed when magnetic nanoparticles are introduced into the biological environment [19,25].

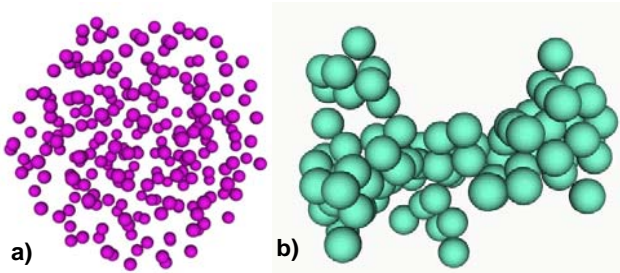

Fig. 7. a) 3D cluster of $N_p = 120$ nanoparticles of the same diameter, b) fractal cluster of $N_p = 90$ nanoparticles with fractal descriptors $D_f = 1.9$, $k_f = 1.7$.

As mentioned above, the effect of the magneto-dipole interaction on the cluster properties can be taken into account by adding the corresponding contribution to the effective magnetic field of the stochastic Landau – Lifshitz equation. Fig. 8a shows the calculated hysteresis loops of dilute assemblies of fractal clusters with fractal descriptors $D_f = 1.9$, $k_f = 1.7$, consisting of nanoparticles of different diameters. The saturation magnetization of the particles and the effective uniaxial anisotropy constant are assumed to be $M_s = 350$ emu/cm$^3$ and $K = 10^5$ erg/cm$^3$, respectively. It is also assumed that the nanoparticles of the cluster are covered with nonmagnetic shells of small thickness, $D_{sh} = 1$ nm.

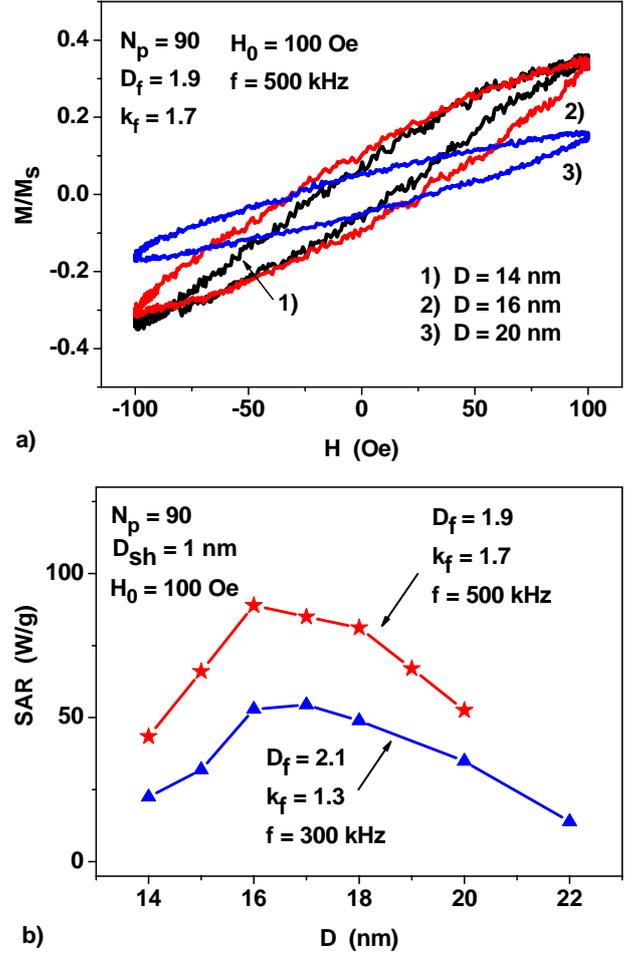

Fig. 8. a) Hysteresis loops of a dilute assembly of fractal clusters of superparamagnetic nanoparticles with uniaxial anisotropy for various particle diameters, b) SAR of dilute assemblies of fractal clusters depending on nanoparticle diameter.

As Fig. 8a shows, the area of the low-frequency hysteresis loops of the assembly remains to depend on the average particle diameter. As a result, the SAR of dilute assemblies of fractal clusters also retains some dependence on the average particle diameter, as Fig. 8b shows. However, comparing Figs. 4b and 8b, it is easy to notice a significant drop in SAR for assemblies of fractal clusters as compared to an assembly of non interacting nanoparticles due to the influence of a strong magnetic dipole interaction between the nanoparticles of a fractal cluster. Nevertheless, a decrease in the SAR of assembly of fractal clusters can be partly avoided [27] if the nanoparticles of the assembly are covered with sufficiently thick nonmagnetic shells, because in this case the intensity of the magnetic dipole interaction between the nearest nanoparticles of the assembly is reduced.

## 5. Conclusions

The results presented above clearly show that the ability of an assembly of magnetic nanoparticles to effectively absorb the energy of external alternating magnetic field significantly depends on the particle size and the type of magnetic anisotropy. According to the calculations performed, for dilute assemblies of iron oxides nanoparticles with effective uniaxial magnetic anisotropy the optimum range of diameters is $D$ = 18 - 22 nm. On the other hand, for quasi-spherical nanoparticles with cubic anisotropy the optimal diameter range shifts to $D$ = 45 - 60 nm. In dense clusters of magnetic nanoparticles there is a sharp decrease in SAR due to the influence of strong magneto- dipole interaction of closest nanoparticles. However, the effect of the magneto-dipole interaction can be weakened by covering the nanoparticles with nonmagnetic shells of sufficient thickness.

Other problems of magnetic hyperthermia which are worth to be studied concern the distribution of heat generated by an assembly of magnetic nanoparticles in a heterogeneous biological environment, with the removal of heat by the blood flows being taken into account. In addition, it is important to estimate the decrease in the amplitude of the alternating magnetic field deep into the biological medium due to the screening of the magnetic field by eddy currents.

**Acknowledgement**

The author gratefully acknowledges the financial support of the Ministry of Science and High Education of the Russian Federation in the framework of Increase Competitiveness Program of NUST «MISIS», contract № K2-2017-008.

———————————————

**References**

1. S. Sun, C. B. Murray, D. Weller, L. Folks and A. Moser, *Science* **287**, 1989 (2000).
2. A. López-Ortega, E. Lottini, C. D. J. Fernández and C. Sangregorio, *Chemistry of Materials*, **27**, 4048 (2015).
3. Q. A. Pankhurst, N. K. T. Thanh, S. K. Jones and J. Dobson, *J. Phys. D: Appl. Phys*. **42**, 224001 (2009).
4. J. Carrey, B. Mehdaoui and M. Respaud, *J. Appl. Phys*. **109**, 083921 (2011).
5. D. Ortega and Q. A. Pankhurst, in *Nanoscience: Nanostructures through Chemistry*, ed. P. O'Brien, The Royal Society of Chemistry, Cambridge, 2012, vol. 1, pp. 60–88.
6. S. Dutz and R. Hergt, *Int. J. Hyperthermia* **29**, 790 (2013).
7. E. A. Périgo, G. Hemery, O. Sandre, D. Ortega, E. Garaio, F. Plazaola, F. J. Teran, *Appl. Phys. Rev*. **2**, 041302 (2015).
8. R. Hergt, *et al*. *J. Magn. Magn. Mater*. **293**, 80 (2005).
9. M. Kallumadil, *et al*. *J. Magn. Magn. Mater*. **321**, 1509 (2009).
10. B. Mehdaoui, *et al*. *J. Magn. Magn. Mater*. **322**, L49 (2010).
11. N.A. Usov, *J. Appl. Phys*. **107**, 123909 (2010).
12. B. Mehdaoui, *et al.. Adv. Funct. Mater*. **21**, 45731 (2011).
13. C. Martinez-Boubeta, *et al. Adv. Funct. Mater*. **22**, 3737 (2012).
14. P. Guardia, *et al. ACS Nano*. **6**, 3080 (2012).
15. S. A. Gudoshnikov, *et al. J. Magn. Magn. Mater*. **324**, 3690 (2012).
16. N.A. Usov and B.Y. Liubimov, *J. Appl. Phys*. **112**, 023901 (2012).
17. D. Serantes, *et al. J. Phys. Chem. C*. **118**, 5927 (2014).
18. R. Di Corato, *et al. Biomaterials*. **35**, 6400 (2014).
19. M. L. Etheridge, *et al. Technology* **2**, 214 (2014).
20. S. Ruta, R. Chantrell and O. Hovorka, *Sci Reports*, **5**, 9090 (2015).
21. M. E. Materia, *et al. Langmuir*. **31**, 808 (2015).
22. C. Blanco-Andujar, *et al. Nanoscale*. **7**, 1768 (2015).
23. I. Conde-Leboran, *et al. J. Phys. Chem. C*. **119**, 15698 (2015).
24. K. Simeonidis, *et al. Sci. Rep*. **6**, 38382 (2016).
25. S. Jeon, *et al. Nanoscale* **8**, 16053 (2016).
26. B. Sanz, *et al. Sci. Rep*. **6**, 38733 (2016).
27. N. A. Usov, O. N. Serebryakova and V. P. Tarasov, *Nanoscale Res. Lett*. **12**, 489 (2017).
28. S. Chikazumi, *Physics of Magnetism* (Wiley, New York, 1964).
29. W.F. Brown, Jr., *Micromagnetics* (Wiley-Interscience, New York - London, 1963).
30. A. Aharoni, *Introduction to the Theory of Ferromagnetism* (Clarendon Press, Oxford, 1996)
31. L. D. Landau and E. M. Lifshitz, *Electrodynamics of Continuous Media* (Pergamon, New York, 1984).
32. W. F. Brown, Jr. and A. H. Morrish, *Phys. Rev*. **105**, 1198 (1957).
33. N. A. Usov and J. M. Barandiarán, *J. Appl. Phys*. **112,** 053915 (2012).
34. N. A. Usov and S. E. Peschany. *J. Magn. Magn. Mater*. **118**, L290 (1993).
35. N. A. Usov and Yu. B. Grebenshchikov, Micromagnetics of small ferromagnetic particles. In *Magnetic Nanoparticles*, edited by S. P. Gubin (Wiley, New York, 2009), Chap. 8.
36. L. Landau and E. Lifshitz, *Phys. Z. Sowjetunion* **8**, 153 (1935).
37. L. Neel, *Ann. Geophys*. **5**, 99 (1949).
38. W. F. Brown, Jr., *Phys. Rev*. **130**, 1677 (1963).
39. I. Klik and L. Gunther, *J. Stat. Phys*. **60**, 473 (1990).
40. D. Weller and A. Moser, *IEEE Trans. Magn*. **35**, 4423 (1999).
41. W. T. Coffey, Yu. P. Kalmykov, and J. T. Waldron, *The Langevin Equation*, 2$^{nd}$ ed. (World Scientific, Singapore, 2004).

42. J.L. Garcia-Palacios and F.J. Lazaro, *Phys. Rev.* B **58**, 14937 (1998).
43. W. Scholz, T. Schrefl and J. Fidler, *J. Magn. Magn. Mater.* 233, 296 (2001).
44. D. V. Berkov, *IEEE Trans. Magn.* 38, 2489 (2002).
45. V. A. Bautin, *et al. J. Magn. Magn. Mater.* **460,** 278 (2018).
46. I. A. Brezovich, *Med. Phys. Monogr.*, **16**, 82 (1988).
47. N. A. Usov, M. S. Nesmeyanov and V. P. Tarasov, *Sci. Reports*, **8**, 1224 (2018).
48. S. R. Forrest and T. A. Witten Jr., *J. Phys. A: Math. Gen.* **12**, L109 (1979).
49. A. V. Filippov, M. Zurita and D. E. Rosner, *J. Colloid. Interface Sci.* **229**, 261 (2000).